\documentclass[graybox]{svmult}

\usepackage{helvet}         
\usepackage{courier}        
\usepackage{type1cm}        
\usepackage{makeidx}         
\usepackage{graphicx}        
\usepackage{multicol}        
\usepackage[bottom]{footmisc}

\usepackage{amsmath,amsfonts,amssymb,mathtools,latexsym}
\usepackage[nosort]{cite}
\usepackage{hyperref}
\usepackage{graphicx}
\usepackage{color}
\usepackage{epsfig}
\usepackage{psfrag}	
\usepackage{tikz}
\usepackage{slashed}
\usepackage{bbold}
\usepackage{tensor}
\usepackage{ulem}
\usepackage[font=small,labelfont=bf,width=1\textwidth]{caption}
\usepackage{comment}
\usepackage[all]{xy}
\usepackage{multirow}
\usepackage{float}
\usepackage{pifont}
\usepackage{cite}
\usepackage{color}

\usepackage{eufrak}
\usepackage{tocloft} 
\usepackage{ulem}
\usepackage{soul,xcolor}
\setstcolor{red}
\usepackage{appendix}
\usepackage{etoolbox}

\definecolor{ZKgreen}{rgb}{0,0.6,0.3}

\DeclareMathOperator{\Tr}{Tr}

\let\Re\undefined
\DeclareMathOperator{\Re}{Re}
\let\Im\undefined
\DeclareMathOperator{\Im}{Im}

\DeclareMathOperator{\vol}{vol}

\newcommand{\bea}{\begin{eqnarray}}
\newcommand{\eea}{\end{eqnarray}}
\newcommand{\beq}{\begin{equation}}
\newcommand{\eeq}{\end{equation}}
\newcommand{\bal}{\begin{equation}\begin{aligned}}
\newcommand{\eal}{\end{aligned} \end{equation}}

\newcommand{\nn}{\nonumber}

\newcommand{\bD}{{\mathbb D}}

\newcommand{\cM}{{\mathcal M}}
\newcommand{\cN}{{\mathcal N}}
\newcommand{\cP}{{\mathcal P}}

\newcommand{\cO}{{\mathcal O}}

\newcommand{\cW}{{\mathcal W}}

\begin{document}

\title*{Integral Identities from Symmetry Breaking of Conformal Defects}
\author{Ziwen Kong}
\institute{Ziwen Kong \at Deutsches Elektronen-Synchrotron DESY, Notkestr. 85, 22607 Hamburg, Germany, \email{zwn.kong@gmail.com}}
\maketitle

\abstract{In conformal field theory, the insertion of a defect breaks part of the global symmetry and gives rise to defect operators such as the tilts and displacements. We establish identities relating the integrated four-point functions of such operators to their two-point functions, derived both from the geometric properties of the defect conformal manifold, which is the symmetry-breaking coset, and from the Lie algebra of the corresponding broken symmetry generators. As an explicit example, we demonstrate these integral identities in the case of the 1/2 BPS Maldacena-Wilson loop in $\cN = 4$ SYM. This contribution serves as a brief review of the main ideas of  Phys. Rev. Lett. 129, 201603 (2022), as well as a short preview of our forthcoming paper with Nadav Drukker and Petr Kravchuk. Here we present an independent derivation of the integral identities that will not appear in that work.}

\section{Introduction}
\label{sec:1}


Defects are ubiquitous in both low- and high-energy physics, appearing in experimental systems as well as theoretical frameworks. In low-energy settings, boundaries and impurities in materials such as ferromagnets or spin chains lead to striking effects, including the Kondo phenomenon. In high-energy theory, line and surface operators, boundary conditions, and interfaces serve as powerful tools for exploring the structure of Quantum Field Theories (QFT). One of the most familiar cases is the Wilson-’t Hooft line operator. 

Many nontrivial conformal defects are also known in conformal field theories (CFT) that preserve a fraction of the bulk symmetry and break the remaining. For instance, a generic conformal defect of dimension $p$ in a $d$-dimensional CFT breaks the translation symmetry in the orthogonal directions, resulting in a modification of the bulk stress-tensor Ward identity by operator-valued contact terms localized on the defect \cite{Billo:2016cpy},
\bal
\label{WardTandD}
\partial^{\mu} T_{\mu \nu} =\delta^{d-p} (x_{\perp}) P^r_{\nu} \bD_{r}\,,
\eal
where $\mu, \nu=1,\cdots,d$ denote bulk spacetime indices, while $r=1,\cdots,d-p$ labels the directions transverse to the defect. The operator $P$ projects bulk indices onto the defect ones. Similarly, if the bulk theory has an internal symmetry $G$ with current $J_{\mu a}$, broken by the defect to $G'$, its Ward identity is modified to
\bal
\label{WardJandt}
\partial^{\mu} J_{\mu a} =\delta^{d-p} (x_{\perp}) P_a^i t_i\,,
\eal
with $i$ an index for the broken internal symmetry generators. Notably, the operator $t_i$, known as the tilt operator, has a conformal dimension $p$ which is equal to that of the defect and is therefore exactly marginal. Within a CFT, exactly marginal operators are special because they permit continuous deformations of the theory, forming what is called the conformal manifold. These operators are typical in supersymmetric theories but rare otherwise. Tilt operators, however, are more general: they originate from internal symmetry breaking by a defect and therefore generally appear in both supersymmetric and non-supersymmetric theories.

Following the same reasoning, if the bulk theory is supersymmetric with supercurrent $j_{\mu \alpha}$, the breaking of supersymmetry gives rise to fermionic operators on the defect, which can be defined as
\bal
\label{WardjandL}
\partial^{\mu} j_{\mu \alpha} =\delta^{d-p} (x_{\perp}) P_{\alpha}^{\sigma} \mathbb{\Lambda}_{\sigma}\,.
\eal
The defect operator $\mathbb{\Lambda}_{\sigma}$ transforms in the spinor representation of $\mathfrak{so} (d-p)$ and under particular representations of the preserved internal symmetry, which is indicated by $\sigma$. Crucially, the $\mathbb{\Lambda}_{\sigma}$ operators are very likely to reside in the same supermultiplets as either the displacement operators or the exactly marginal ones---a fact proven for line defects in \cite{Agmon:2020pde}. For this reason, we will mainly concentrate on integral identities for tilt and displacement operators, with those of $\mathbb{\Lambda}_{\sigma}$ following from supersymmetry.

The plan of this contribution is as follows. In Section \ref{sec:2}, we describe the defect conformal manifold, which is isomorphic to the symmetry-breaking coset. The Zamolodchikov metric is captured by the two-point functions of the tilts, and the Riemann tensor can be expressed as an integrated four-point function, providing an exact relation to the curvature of the coset space. In Section \ref{sec:3}, we present an alternative interpretation of these integral identities via the Lie algebra of the broken internal symmetry generators. We then extend this approach to displacement operators, where we find three integral identities governed by the Lie algebra of the broken conformal symmetry generators. In Section \ref{sec:4}, we demonstrate these identities on previously obtained four-point functions for insertions into the 1/2 BPS Wilson loop in $\cN = 4$ super Yang-Mills theory. Section \ref{sec:5} summarizes the main results and briefly covers additional cases where the integral identities can be used.

\section{Defect Conformal Manifold}
\label{sec:2}

At any point on the defect conformal manifold, which we take as the origin of a coordinate system $w^i$, the defect exactly marginal operators induce infinitesimal deformations to a neighboring point, where correlation functions of any defect operators $\cO$ in the deformed theory can be expressed by 
\bal
\label{tdeformation}
\langle \cO_{\alpha} \cO_{\beta}\cdots \rangle_{w^i} =\langle e^{-\int d^p \tau\, w^i t_i}\cO_{\alpha} \cO_{\beta} \cdots \rangle_{0}\,,
\eal
regardless of whether an action exists. Here the brackets $\langle \cdots \rangle_0$ are to be understood as correlation functions evaluated at the origin of the defect conformal manifold, normalized by the expectation value of the defect $\cW$ without insertions, namely $\frac{\langle \cW[\cdots] \rangle_0}{\langle \cW \rangle}$. It should be noted that we are using $t_i$ for the defect exactly marginal operators, the same notation used for the tilts in \eqref{WardJandt}, but this is not fully rigorous. There might be other exactly marginal operators that are not tilts. However, for simplicity we just ignore the other possibilities.

A defect conformal manifold admits a natural Riemannian structure given by the Zamolodchikov metric
\bal
\label{Zmetric}
g_{ij}=\langle t_i (0) t_j (\infty) \rangle=C_t \delta_{ij}\,,
\eal
where $t (\infty) =\lim\limits_{\tau\rightarrow \infty} \tau^{2p} \, t(\tau)$. Importantly, the metric is defined only locally, meaning it does not reflect the global geometry of the defect conformal manifold. To go beyond the flat-space approximation, we recall that the Riemann tensor of a general vector bundle over the manifold takes the form
\bal
R_{ij\alpha}{}^{\beta} |\cO_{\beta}\rangle=[\nabla_i, \nabla_j] |\cO_{\alpha} \rangle\,,
\eal
where $|\cO_{\alpha} \rangle$ is a vector. For a defect operator $\cO_{\alpha} (0)$,
the two-point function normalization $\langle \cO_{\alpha} (0) \cO_{\beta} (\infty)\rangle =C_{\cO} \delta_{\alpha \beta}$ allows us to lower the index $\beta$ as
\bal
\label{doubleintegratedcurvature}
R_{ij\alpha\beta}=\langle \cO_{\beta}(\infty)|[\nabla_i, \nabla_j] |\cO_{\alpha} (0)\rangle\,.
\eal
The correlators on the defect conformal manifold are given by \eqref{tdeformation}, and the derivatives give insertions of tilt operators leading to the result \cite{Kutasov:1989, Papadodimas:2009eu}
\bal
\label{doubleint}
R_{ij\alpha\beta}= \int d^p \tau_1 d^p \tau_2 \langle t_{[i}(\tau_1) t_{j]}(\tau_2) \cO_{\alpha} (0) \cO_{\beta}(\infty)\rangle_c\, ,
\eal
where $\langle \cdots \rangle_c$ indicates the connected correlator. As shown in \cite{Papadodimas:2009eu, deBoer:2008ss}, in general such an expression for the curvature tensor does not vanish. One way to see this is to adopt a hard-sphere (point-splitting) cutoff as a regularization scheme, which effectively dresses the integration region in \eqref{doubleint} to \cite{Friedan:2012},
\bal
\Sigma_{\varepsilon}&(\tau_1,\tau_2) =\{(\tau_1,\tau_2)|\, \varepsilon^{-1}>|\tau_1|>\varepsilon,\,\varepsilon^{-1}> |\tau_2|>\varepsilon,\, |\tau_1-\tau_2|>\varepsilon \}\,.
\eal

As the 4-point function depends only on the cross-ratio, one can perform one of the two integrals in the curvature \eqref{doubleint} explicitly, reducing the formula to
\bal
\label{renorR}
R_{ij\alpha\beta}= \int d^p \tau F_{\varepsilon}(\tau) \langle t_{i}(1) t_{j]}(\tau) \cO_{\alpha} (0) \cO_{\beta}(\infty)\rangle_c\, ,
\eal
where
\bal
F_{\varepsilon}(\tau)=\left(\int_{ \Sigma_{\varepsilon}(x,x\tau)} -\int_{ \Sigma_{\varepsilon}(x\tau,x)} \right)\frac{dx}{x}\,.
\eal
As was proven in \cite{Friedan:2012},
\bal
\lim\limits_{\varepsilon \rightarrow 0} F_{\varepsilon}(\tau)=\vol_{S^{p-1}} \log|\tau|\,.
\eal
Thus far, our analysis allows us to explicitly write down the practical form of the integrals entering the curvature tensor \eqref{doubleint} \cite{Friedan:2012, Balthazar:2022hzb}. For $p=1$, it is given by
\bal
\label{1dR}
R_{ij\alpha \beta}= 2 \int_{-\infty}^{+\infty} d\tau \, \log|\tau| \langle t_i (1) t_j(\tau) \cO_{\alpha} (0) \cO_{\beta} (\infty) \rangle_c\,,
\eal
and for general $p$, since the four points can always be placed within the same plane by conformal transformation, we can reduce \eqref{doubleint} to an 2d integral,
\bal
R_{ij\alpha\beta}=\frac{1}{2} \vol_{S^{p-1}}\vol_{S^{p-2}} \int d^2\tau (\Im \tau)^{p-2} \log|\tau| \langle t_i (1) t_j(\tau) \cO_{\alpha} (0) \cO_{\beta} (\infty) \rangle_c\,. 
\eal

As mentioned earlier, if all exactly marginal operators result from internal symmetry breaking, the defect conformal manifold is the coset $G/G'$. This is not manifest in the correlation function analysis above, which is local on the defect conformal manifold, but rather a nontrivial statement that allows us to predict the form of the curvature. An illustrative example is the 1/2 BPS Wilson loops in $\cN=4$ super Yang–Mills theory, which break the bulk R-symmetry group $SO(6)$ down to its subgroup $SO(5)$, yielding the coset $SO(6)/SO(5)=S^5$. A more detailed discussion will be given in Section \ref{sec:4}. The curvature of the 5-sphere with radius $r$ is well known and takes the form $R_{ijkl}=r^2 (\delta_{ik} \delta_{jl} -\delta_{il} \delta_{jk})$. Moreover, the radius $r$ can be extended to represent the scale of the coset, as set by the normalization of the tilts $C_t$ in \eqref{Zmetric}. Since $C_t$ appears both in the metric and the curvature tensor, equation \eqref{doubleintegratedcurvature} constitutes a nontrivial relation for integrated correlators.

\section{Algebraic Approach}
\label{sec:3}

In the CFT, integrating the modified Ward identities \eqref{WardJandt} over the whole bulk spacetime leads to commutation relations involving the defect $\cW$ and the broken internal symmetry charges $Q_{i} \equiv \int dS^{\mu} J_{\mu i}$, where $S^{\mu}$ denotes the area element on some hypersurfaces,
\bal
\label{chargeJ}
[Q_{i}, \cW] =\int d^p \tau \, \cW[t_i(\tau)]\,.
\eal
Accordingly, when two broken charges act on a defect with operators $\cO$ inserted at $0$ and $\infty$, their commutator produces
\bal
\label{commutatorJ}
\langle \cO_{\beta}(\infty)|[Q_i, Q_j] |\cO_{\alpha} (0)\rangle =\int d^p \tau_1 d^p \tau_2 \langle t_{[i}(\tau_1) t_{j]}(\tau_2) \cO_{\alpha} (0) \cO_{\beta}(\infty)\rangle_c\, ,
\eal
which reproduces the same integral as in \eqref{doubleint}. This offers an alternative explanation for the integral identities obtained earlier. Here, instead of being treated as bundles on the defect conformal manifold, the operators $\cO$ are interpreted as charged objects transforming in finite-dimensional representations of the internal symmetry group. It should be stressed, however, that the two viewpoints are equivalent, as the geometry of the bundles associated with these operators is entirely fixed by group-theoretic data \cite{deBoer:2008ss}.

Still, the algebraic perspective is very suggestive and can be straightforwardly applied to other forms of symmetry breaking, including conformal and supersymmetric cases, as highlighted in the introduction section \ref{sec:1}. As an illustration, recall that in Euclidean signature the conformal Lie algebra is generated by $\{ P_{\mu} ,M_{\mu \nu}, K_{\mu} ,D\}$, satisfying the commutation relations:
\bal
\label{conformalalgebra}
{}[P_\mu,K_\nu]&=-(2\delta_{\mu\nu}D+2M_{\mu\nu})\,,
\;&
[D,P_\mu]&=P_\mu\,,
\;&
[D,K_\mu]&=-K_\mu\,,
\\
[M_{\mu\nu},M_{\rho\sigma}]&=\delta_{\mu[\rho}M_{\sigma]\nu}-\delta_{\nu[\rho}M_{\sigma]\mu}\,,
\;&
[P_\mu,M_{\nu\rho}]&=\delta_{\mu[\nu}P_{\rho]}\,,
\;&
[K_\mu,M_{\nu\rho}]&=\delta_{\mu[\nu}K_{\rho]}\,.
\eal
and with all other brackets vanishing. Since the transverse translation, rotation and special conformal transformation charges are defined by the stress-tensor in the usual way as $P_{\mu}\equiv \int dS^{\rho} T_{\mu \rho}$, $M_{\mu \nu} \equiv \int dS^{\rho} x_{[\mu}T_{\nu]\rho}$ and $K_{\mu}\equiv \int dS^{\rho} x^{\nu}x_{\nu} T_{\mu \rho}$, analogous to the case of broken internal symmetries, integrating the modified Ward identities \eqref{WardTandD} now yields the broken charges $\{P_r,M_{u r},K_{r}\}$ and their commutators with the defect
\bal
[P_r, \cW]=&\int d^p \tau\, \cW[\bD_r (\tau)]\,,\\
[M_{u r}, \cW]=&\int d^p \tau\, \tau_u \cW[\bD_r (\tau)]\,,\\
[K_r, \cW]=&\int d^p \tau\, \tau^u \tau_u \cW[\bD_r (\tau)]\,,
\eal
with $u=1,\cdots,p$ are directions along the defect. Following the same logic as the broken internal symmetries, for $\mathfrak{g}_r\in \{P_r,M_{u r},K_{r}\}$, acting twice on the defect and forming the commutator results in
\bal
\label{doubleintD}
{}&\langle \cO_{\beta}(\infty)|[\mathfrak{g}_r, \mathfrak{g}_s] |\cO_{\alpha} (0)\rangle \\
=&\int d^p \tau_1 d^p \tau_2 \left(\cM_{\mathfrak{g}_r}(\tau_1) \cM_{\mathfrak{g}_s}(\tau_2) \langle \bD_{r}(\tau_1) \bD_{s}(\tau_2) \cO_{\alpha} (0) \cO_{\beta}(\infty)\rangle_c -r\leftrightarrow s \right)\, ,
\eal
where $\cM_{\mathfrak{g}}(\tau)$ is a monomial in $\tau$ that depends on the choice of $\mathfrak{g}$. Explicitly, $\cM_{P_r}=1$, $\cM_{M_{ur}}=\tau_u$, and $\cM_{K_r}=\tau^u \tau_u$. 

We now examine three cases in which the left-hand side of \eqref{doubleintD} does not vanish in general, namely $[P_r,K_s]$, $[M_{ur},M_{vs}]$ and $[K_r,P_s]$. Let us begin with $[P_r,K_s]$,
\bal
\langle \cO_{\beta}(\infty)|[P_r, K_s] |\cO_{\alpha} (0)\rangle =\!\!\!\!\!\!\!\!
\int\limits_{\Sigma_{\varepsilon}(\tau_1,\tau_2)} \!\!\!\!\!\!\!\!
 d^p \tau_1 d^p \tau_2 \big(&\tau_2^2 \langle \bD_{r}(\tau_1) \bD_{s}(\tau_2) \cO_{\alpha} (0) \cO_{\beta}(\infty)\rangle_c\\
 -\tau_1^2 &\langle \bD_{r}(\tau_2) \bD_{s}(\tau_1) \cO_{\alpha} (0) \cO_{\beta}(\infty)\rangle_c \big)\, ,
\eal
Applying the same renormalization procedure as in \eqref{renorR}, the double integrals above reduce to a single cross-ratio integral, given explicitly by
\bal
{}&\lim\limits_{\varepsilon\rightarrow 0}\int d^p \tau F_{\varepsilon}(\tau) \tau^2 \langle \bD_{r}(1) \bD_{s}(\tau) \cO_{\alpha} (0) \cO_{\beta}(\infty)\rangle_c \\
=&\vol_{S^{p-1}} \int d^p \tau \, \tau^2 \log|\tau| \langle \bD_{r}(1) \bD_{s}(\tau) \cO_{\alpha} (0) \cO_{\beta}(\infty)\rangle_c\,.
\eal
In the same manner, we can find the reduced integrals corresponding to the other two commutators. Just as with the tilt operators, these expressions can be simplified based on the dimension of the defect. When $p=1$,
\bal
\label{intD1d}
\int & d \tau\, \log|\tau| \langle \bD_{r}(1) \bD_{s}(\tau) \cO_{\alpha} (0) \cO_{\beta}(\infty)\rangle_c\\
&\qquad \qquad \qquad \;\; = \delta_{rs} \Delta_{\cO} \langle \cO_{\alpha} (0) \cO_{\beta}(\infty)\rangle -\langle [M_{rs},\cO_{\alpha}] (0) \cO_{\beta}(\infty)\rangle\,,\\
\int & d \tau\,\tau \log|\tau| \langle \bD_{r}(1) \bD_{s}(\tau) \cO_{\alpha} (0) \cO_{\beta}(\infty)\rangle_c= \frac{1}{2}\langle [M_{rs},\cO_{\alpha}] (0) \cO_{\beta}(\infty)\rangle\,,\\
\int & d \tau\, \tau^2 \log|\tau| \langle \bD_{r}(1) \bD_{s}(\tau) \cO_{\alpha} (0) \cO_{\beta}(\infty)\rangle_c\\
&\qquad \qquad \quad \;\;= -\delta_{rs} \Delta_{\cO} \langle \cO_{\alpha} (0) \cO_{\beta}(\infty)\rangle -\langle [M_{rs},\cO_{\alpha}] (0) \cO_{\beta}(\infty)\rangle\,.
\eal
These identities also appear in \cite{Gabai:2025zcs, Gabai:2025hwf}. For general $p$,
\bal
\frac{1}{4} \vol_{S^{p-1}} \vol_{S^{p-2}} \!\int & d^2 \tau\, (\Im \tau)^{p-2}\log|\tau| \langle \bD_{r}(1) \bD_{s}(\tau) \cO_{\alpha} (0) \cO_{\beta}(\infty)\rangle_c\\
& = \delta_{rs} \Delta_{\cO} \langle \cO_{\alpha} (0) \cO_{\beta}(\infty)\rangle -\langle [M_{rs},\cO_{\alpha}] (0) \cO_{\beta}(\infty)\rangle\,,\\
\frac{1}{4} \vol_{S^{p-1}} \vol_{S^{p-2}} \!\int & d^2 \tau\,(\Im \tau)^{p-2} \Re \tau \log|\tau| \langle \bD_{r}(1) \bD_{s}(\tau) \cO_{\alpha} (0) \cO_{\beta}(\infty)\rangle_c\\
&\qquad \qquad \qquad \qquad \quad \quad\;= \frac{p}{2}\langle [M_{rs},\cO_{\alpha}] (0) \cO_{\beta}(\infty)\rangle\,,\\
\frac{1}{4} \vol_{S^{p-1}} \vol_{S^{p-2}} \!\int & d^2 \tau\, (\Im \tau)^{p-2} |\tau|^2 \log|\tau| \langle \bD_{r}(1) \bD_{s}(\tau) \cO_{\alpha} (0) \cO_{\beta}(\infty)\rangle_c\\
&\!\! = -\delta_{rs} \Delta_{\cO} \langle \cO_{\alpha} (0) \cO_{\beta}(\infty)\rangle -\langle [M_{rs},\cO_{\alpha}] (0) \cO_{\beta}(\infty)\rangle\,.
\eal
Another derivation of these integral identities will appear in our upcoming paper \cite{upcoming}.

We must emphasize that although two of the identities come from $[K,P]$ and $[P,K]$, which are opposite to one another, this does not make them redundant---the identities remain independent. In fact, as we prove in \cite{upcoming}, even though there are other commutators in \eqref{conformalalgebra}, the three identities given above already exhaust all renormalization scheme–independent, non-vanishing relations for displacement operators. For the tilt operators, by comparison, \eqref{commutatorJ} provides the unique such identity. Note that in this contribution we do not consider anomalies; a detailed discussion of anomalies and the additional integral identities they give rise to will appear in \cite{upcoming}.

\section{Example: 1/2 BPS Wilson loops in $\cN=4$ super Yang-Mills theory}
\label{sec:4}

In this section, we demonstrate the integral identities derived in the previous sections using the example of the 1/2 BPS Wilson loop in $\cN=4$ super Yang-Mills theory \cite{Maldacena:1998im, Drukker:1999zq}. We begin by recalling some relevant facts about the theory. Consider a 1/2 BPS Wilson loop along the Euclidean time direction, defined as
\bal
\label{WL}
W=\Tr\cP e^{\int(iA_0+\Phi_6)dt}\,,
\eal
it is a line defect, hence $p=1$. The defect CFT perspective on this observable has been developed in \cite{Correa:2012, Drukker:2012, Correa:20122, Gromov:2012}. There are six scalar fields $\Phi_I$ in the theory. Of these, $\Phi_6$ is marginally irrelevant, while the remaining five scalars are tilt operators appearing in \eqref{WardJandt} \cite{Alday:2007, Polchinski:2011, Beccaria:2017, Bruser:2018, Cuomo:2021, Giombi:2017}. For consistency with the previous sections, we will continue to denote these five tilts in the theory by $t_i$. The displacement is gotten by differentiating with respect to a normal direction and is the insertion of $\bD_r=iF_{r\tau}+D_r\Phi_6$. Together with the fermionic operator $\mathbb{\Lambda}_{\sigma}$ in \eqref{WardjandL}, these operators form the displacement supermultiplet,
\bal
t_i \rightarrow \mathbb{\Lambda}_{\sigma} \rightarrow \bD_{r}\,.
\eal
It follows that the two-point functions of the tilts and of the displacement operators are related by \cite{Correa:20122, Bianchi:2018zpb}
\begin{equation}
\label{N=4-brems}
C_t=\frac{C_\bD}{6}
=\frac{\sqrt\lambda}{2\pi^2}\frac{I_2(\sqrt{\lambda})}{I_1(\sqrt{\lambda})}\,,
\end{equation}
where $\lambda$ is the 't Hooft coupling and $I_n$ are modified
Bessel functions. Similarly to $C_t$ in \eqref{Zmetric}, $C_{\bD}$
is defined as $ C_{\bD}\delta_{rs}=\langle \bD_{r} (0) \bD_s(\infty)\rangle$. At weak and strong coupling, they are
\begin{equation}
\label{CPhi-expand}
C_t=
\begin{cases}
\frac{\lambda}{8\pi^2}- \frac{\lambda^2}{192\pi^2}+\frac{\lambda^3}{3072\pi^2}
-\frac{\lambda^4}{46080\pi^2}
+O(\lambda^5)\,,\\
\frac{\sqrt{\lambda}}{2\pi^2} -\frac{3}{4\pi^2} +\frac{3}{16\pi^2\sqrt{\lambda}}+\frac{3}{16\pi^2\lambda}+O(\lambda^{-3/2})\,,
\end{cases}
\end{equation}
and $C_{\bD}$ can then be readily read off \eqref{N=4-brems}.

Four-point functions of operators in the displacement supermultiplet are well understood (see \cite{Beccaria:2017, Liendo:1806, Ferrero:2103, Ferrero:2023znz, Ferrero:2023gnu, Cavaglia:2022qpg}). For the tilt operators
\bal
\label{eq:N=4Gs}
\langle t_k (0)t_j(\tau) t_i(1) t_l(\infty)\rangle
=\frac{C_t^2}{\tau^2} \big(\delta_{jk}\delta_{il}G_1(\tau) +\delta_{ij]}\delta_{kl} G_2(\tau) +\delta_{ik}\delta_{jl} G_3(\tau) \big)\,,
\eal
with
\bal
\label{ftoG}
G_1= \tau^2\left(\frac{\tau-1}{\tau^2}f(\tau)\right)',
\;
G_2=-\tau^2 \left(\frac{f(\tau)}{\tau}\right)',
\;
G_3=\tau^2 \mathsf{f} -\tau^2 \left(\frac{\tau-1}{\tau}f(\tau)\right)',
\eal
where $\mathsf{f}=3\frac{I_1(\sqrt\lambda)I_3(\sqrt\lambda)}{I_2(\sqrt\lambda)^2}$ is the so-called topological constant, determined from the topological sector of the correlators \cite{Liendo:1806, Drukker:1806, Giombi:1802}. We also give its weak- and strong-coupling expansions,
\bal
\mathsf{f}= \begin{cases}
2+\frac{\lambda}{24}-\frac{\lambda^2}{480}+\frac{7\lambda^3}{69120}
-\frac{11\lambda^4}{2322432}+O(\lambda^5)\,,\\
3-\frac{3}{\sqrt\lambda}+\frac{45}{8\lambda^{3/2}}
+\frac{45}{4\lambda^2}+\frac{1215}{128\lambda^{5/2}}+O(\lambda^{-3})\,.
\end{cases}
\eal
As discussed earlier, since the displacement operators are related to the tilts by supersymmetry, their four-point functions can also be written in terms of the topological constant $\mathrm{f}$, the function $f(\tau)$, and its higher derivatives with respect to $\tau$,
\bal
\label{4displacements}
\langle \bD_{p}(0) \bD_{s} (\tau) \bD_{r} (1) \bD_{q} (\infty) \rangle
=\frac{C_{\bD}^2}{\tau^4 } (\delta_{sp} \delta_{r q} H_1(\tau) 
+\delta_{r s} \delta_{pq} H_2(\tau) 
+\delta_{r p} \delta_{s q} H_3(\tau))\,,
\eal
where
\begin{align}
\label{Hi}
H_1(\tau)
&=-\frac{\tau^2}{36}\partial_\tau\bigg(\!\tau^2\partial_\tau\bigg(\! \tau^2\partial_\tau
\bigg(\! \frac{\tau}{(1-\tau)^2}\partial_\tau\bigg(\!(1-\tau)^6
\partial_\tau\bigg(\!\frac{f(\tau)}{(1-\tau)\tau^3}\bigg)\!\bigg)\!\bigg)\! \bigg)\! \bigg)\,,
\\
H_2(\tau)
&=-\frac{\tau^2}{36}\partial_\tau\bigg(\!\tau^2\partial_\tau\bigg(\!\tau^2\partial_\tau
\bigg(\!\tau(1-\tau)^3\partial_\tau\bigg(\!\frac{1}{\tau^4}
\partial_\tau\bigg(\!\frac{\tau^2f(\tau)}{1-\tau}\bigg)\!\bigg)\!\bigg)\!\bigg)\!\bigg)\,,
\\
H_3(\tau)
&=\tau^4\mathsf{f}
+\frac{\tau^2}{36}\partial_\tau\bigg(\!\tau^2\partial_\tau\bigg(\!\tau^2\partial_\tau\bigg(\!\frac{\tau}{(1-\tau)^2}\partial_\tau\bigg(\!\frac{(1-\tau)^6}{\tau^4}\partial_\tau
\bigg(\!\frac{\tau^2f(\tau)}{1-\tau}\bigg)\!\bigg)\!\bigg)\!\bigg)\!\bigg)\,.
\end{align}
Using the knowledge gathered above, we employ these tilt and displacement four-point functions to demonstrate the integral identities obtained in the previous sections \ref{sec:2} and \ref{sec:3} at strong coupling limit. For brevity, we avoid the weak-coupling analysis, which is complicated by contact terms. This issue is, however, fully addressed; for details see \cite{Gabai:2025zcs, upcoming, Cavaglia:2022qpg}.

As we mentioned at the end of section \ref{sec:2}, the defect conformal manifold in this case has the geometry of the coset $S^5=SO(6)/SO(5)$ with radius $C_t^{1/2}$ \eqref{Zmetric}, and hence its curvature is
\bal
\label{curvature5sphere}
R_{ijkl}=C_t (\delta_{ik} \delta_{jl} -\delta_{il} \delta_{jk})\,,
\eal
which we would now like to identify with the integrated four-point functions in \eqref{1dR} for $\cO_{\alpha}=t_k$ and $\cO_{\beta}=t_l$. In the algebraic interpretation of this integral identity, for a generator $J_{IJ} \in SO(6)$, its commutator satisfies
\bal
[J_{IJ},J_{KL}]=-\delta_{IK} J_{JL} +\delta_{JK} J_{IL} -\delta_{JL} J_{IK} +\delta_{IL} J_{JK}\,.
\eal
In this case, the broken charges defined in \eqref{chargeJ} are $Q_i\equiv J_{i6}$ with algebra
\bal
[Q_i,Q_j]= -J_{ij}\,,
\eal
Given that the tilts $t_i$ transform in the fundamental representation of $SO(5)$, the left-hand side of equation \eqref{commutatorJ} accordingly takes the form
\bal
\langle t_l(\infty)|[Q_i, Q_j] |t_k (0)\rangle =C_t(\delta_{ik} \delta_{jl} -\delta_{il} \delta_{jk})\,,
\eal
which matches \eqref{curvature5sphere}. This is an explicit example showing that the geometric and algebraic approaches are consistent. In any case, the resulting identity is
\bal
\label{tiltint}
2 \int_{-\infty}^{+\infty} d\tau \, \log|\tau| \langle t_i (1) t_j(\tau) t_k (0) t_l (\infty) \rangle=C_t(\delta_{ik} \delta_{jl} -\delta_{il} \delta_{jk})\,.
\eal
Plugging in the tilt four-point function \eqref{eq:N=4Gs} and using the crossing relations
\bal
\tau^2 G_1(1-\tau)=(1-\tau)^2 G_2(\tau)\,,\qquad \tau^2 G_3(1-\tau)=(1-\tau)^2 G_3(\tau)\,,
\eal
it turns out that the integral appearing on the left-hand side of \eqref{tiltint} can be expressed as an integral over $[0,1]$,
\bal
2C_t^2 (\delta_{ik} \delta_{jl} -\delta_{il} \delta_{jk}) \int_0^1 \frac{d\tau}{\tau^2} \log \tau (G_3(\tau)+G_2(\tau)-2G_1(\tau))\,.
\eal
Comparing with the right-hand side of \eqref{tiltint}, it exhibits the same tensor structure. Consequently, the integral identity reduces to a relation between the integrated $G_{1,2,3}(\tau)$ functions and the tilt normalization $C_t$. The fact that \eqref{tiltint} is satisfied indicates that the strong-coupling expansion of $f(\tau)$ up to fourth order \cite{Beccaria:2017, Liendo:1806, Ferrero:2103, Ferrero:2023znz, Ferrero:2023gnu} is consistent.

As for the conformal symmetry breaking, the 1/2 BPS Wilson loop breaks the bulk conformal group in Euclidean signature $SO(d+1,1)$ to $SL(2,\mathbb{R}) \times SO(d-1)$, namely the one-dimensional conformal group times the rotations around the line. We may recast the displacement operator identities in \eqref{intD1d} into one unified expression. For $\cO_{\alpha}=\bD_k$ and $\cO_{\beta}=\bD_l$, since the dimensions of the displacement operators and the action of $M_{rs}$ on them are known, \eqref{intD1d} can be rewritten as
\bal
\label{DDDD}
\int_{-\infty}^\infty d\tau \log|\tau| \tau^{n}
\langle \bD_{p}(0) \bD_{s}(\tau) \bD_{r}(1) \bD_{q}(\infty)\rangle_c
=\delta_{sp}\delta_{rq }Y^{n}_1+\delta_{r s} \delta_{pq} Y^{n}_2+\delta_{r p}\delta_{sq}Y^{n}_3\,,
\eal
with $n=0,1,2$, and
\begin{align}
\label{Ys}
Y^0_1&=C_\bD\,,\quad
& Y^0_{2}&=-2C_\bD,\quad
& Y^0_3&=-C_\bD\,,\nn\\
Y^1_1&=-C_\bD/2\,,\quad
& Y^1_{2}&=0\,,\quad
&Y^1_3&=C_\bD/2\,,\\
Y^2_1&=C_\bD\,,\quad
& Y^2_{2}&=2C_\bD\,,\quad
& Y^2_3&=-C_\bD.\nn
\end{align}
We checked that there are only three independent components---for example, $Y_3^n$---with all others related to them via the one-dimensional conformal symmetry. Following the same procedure as for tilts, substituting the displacement four-point functions into the identities provides a nontrivial check on the strong-coupling $f(\tau)$ up to fourth order.

\section{Conclusion}
\label{sec:5}

Modified Ward identities of conserved currents in the presence of a defect induce a distinguished set of conformal primary operators on the defect. This set universally includes the displacement operator, associated with broken conformal symmetry, and also the tilt operators, tied to internal symmetry breaking. Our analysis reveals identities between the integrated four-point functions and two-point functions of these defect operators. These can be understood from two perspectives: the geometry of the defect conformal manifold, described by the symmetry-breaking coset, or the Lie algebra of the broken symmetry generators. The geometric viewpoint is particularly well suited for the tilt operators, since being exactly marginal they can be identified with tangent vectors on the defect conformal manifold. In this framework, their two-point functions define the Zamolodchikov metric, while their integrated four-point functions determine the Riemann tensor. The algebraic perspective, however, is broader: it enables us to generalize the construction to displacement operators, whose integral identities are dictated by the commutators of the broken conformal symmetry generators. Eventually, we find that the displacement four-point function admits three renormalization scheme-independent, non-vanishing integral identities, whereas for the tilts there is only one. It is worth noting that, in addition to the non-vanishing identities, there also exist integrated correlators that always vanish, yet still impose non-trivial constraints on the defect CFT \cite{Gabai:2025zcs, upcoming}.

The integral identities derived in this work are very general and apply to any conformal defect. Moreover, they are non-perturbative: for example, in the case of 1/2 BPS Wilson loops in super Yang-Mills theory, we have demonstrated these identities up to fourth order in the strong coupling expansion using previously known four-point functions. However, this is not a limitation---the identities hold at all orders, for any value of $\lambda$, finite or not. As detailed in \cite{upcoming}, these integral identities have broad applicability. Examples include 1/2 BPS Wilson loops in ABJM theory \cite{Drukker:0912, Bianchi:2004}, the magnetic lines in the $O(N)$ model \cite{Cuomo:2112, Gimenez-Grau:2208, Belton:2025hbu}, 1/2 BPS line defects in the holographic dual of type IIB string theory on $AdS_3\times S_3\times T_4$ \cite{Bliard:2024bcz}, 1/2 BPS surface operators in 6d $\cN=(2,0)$ theory \cite{Drukker:2020swu}, and various line defects in 3d super Chern-Simons theory \cite{Drukker:2022txy, Pozzi:2024xnu}.

\begin{acknowledgement}
We are grateful to N. Drukker and P. Kravchuk for their collaborative work in \cite{upcoming}, from which some results are presented in this contribution. We especially thank N. Drukker for carefully reviewing the draft and providing valuable suggestions. Z.K. is supported by ERC-2021-CoG---BrokenSymmetries 101044226.
\end{acknowledgement}
%


\end{document}